# Do Dark Gravity Theories Predict Opera Superluminal Neutrinos and LENR Phenomena?

F Henry-Couannier
Centre de Physique des Particules de Marseille
July 1, 2012

**Abstract**

We investigate whether Dark Gravity theories (DG) with two conjugate metrics $g_{\mu\nu}$ and $\bar{g}_{\mu\nu} = \eta_{\mu\rho} \eta_{\nu\lambda} g^{\rho\lambda}$ where $\eta_{\mu\rho}$ is supposed to be a background non dynamical and flat metric or an auxiliary field, actually predicted the occurrence of apparently superluminal propagations (from our metric side $g_{\mu\nu}$ point of view) such as the one recently reported by the Opera experiment. We find that indeed such theories could predict the order of magnitude of the superluminal velocity and even explain the apparent conflict with the SN1987 normal neutrino speeds provided the neutrinos are able to oscillate between the two conjugate metrics while propagating in a dense medium. We then explain the theoretical motivations and explore all possible phenomenological consequences of the field discontinuities naturally expected in some Dark Gravity theories. Since the Opera result was not confirmed, these discontinuities do not actually allow a propagation of neutrinos oscillating between the two conjugate metrics.

Two attempted crosslists (gr-qc, hep-th) for this article were rejected « upon a notice from Arxiv moderators, who determined the submission to be inappropriate for the gr-qc (General Relativity and Quantum Cosmology) and hep-th (High Energy Physics-Theory) subject classifications. »So if you think that this article and theoretical works in the references deserve to be known by theorists, please make it known.

## I Introduction

The Opera collaboration has recently published [1] the measurement of a superluminal velocity of muon neutrinos. Indeed, the 17.5 GeV (this is a mean) neutrinos propagating underground from CERN to the Gran Sasso Laboratory 730 km away had a measured travel time 60 ns less than expected for ultra-relativistic particles suggesting that neutrinos were propagating at a speed $v_\nu$ slightly higher than the speed of light c :

$$\delta v_\nu / c = (v_\nu - c)/c = (2.48 \pm 0.28 \text{ (stat.)} \pm 0.30 \text{ (sys.)}) \times 10^{-5}$$

The result is compatible with an earlier less sensitive measurement by the MINOS [2] experiment which result appeared to deviate from c by only 1.8 standard deviations but the result also severely conflicts with the observations of 10 MeV neutrinos from SN1987 ($\delta v_\nu / c < 2 \cdot 10^{-9}$) [3] a noticeable difference being that the latter mainly propagated in vacuum.

If the result is confirmed, it has to be stressed that it would not necessarily imply that the Opera neutrinos was genuine tachyons propagating in our metric $g_{\mu\nu}$ but instead might have propagated in another gravitational field which metric would be, let say, $\bar{g}_{\mu\nu}$ at a speed smaller then the local speed of light, still c, in such metric, but resulting in a higher speed of light from our point of view, since we are living in $g_{\mu\nu}$ where our reference clocks and rods are affected differently than in $\bar{g}_{\mu\nu}$. This is exactly the same idea as the one recently advocated by J. W. Moffat to also explain the Opera superluminal velocities [4]. It is also appropriate to recall that even in GR the velocity of light or any ultra-relativistic particle propagating in a gravitational field different from the one our rods and clocks feel on earth, for instance in the vicinity of a far away compact object, as measured with respect to our local rods and clocks also appears subluminous or superluminous as explained in more details in [5] although the locally measured speed of light is everywhere still of course c. The interesting new phenomenology allowed by $\bar{g}_{\mu\nu}$ is that the particles need not propagate in another distant gravitational field (necessarily far away to be very different from our local one) but just here and now in $\bar{g}_{\mu\nu}$.

## II Superluminal propagations: from $\bar{g}_{\mu\nu}$ to $g_{\mu\nu}$

Three somewhat different approaches [6][7][8] led to theories with two conjugate metrics linked by a relation which is perfectly or approximately $\bar{g}_{\mu\nu} = \eta_{\mu\rho} \eta_{\nu\lambda} g^{\rho\lambda}$ where $\eta_{\mu\rho}$ is here as in [6] supposed to be a background non dynamical Minkowkian metric although the same kind of relation were derived by Hossenfelder [8] starting from a theoretical framework where $\eta_{\mu\rho}$ is a truly dynamical field. In such theories, in principle, fields propagating in $\bar{g}_{\mu\nu}$ cannot interact through electromagnetic, weak or strong interactions with fields in $g_{\mu\nu}$ so we have to postulate another mechanism that would in certain cases allow the wave packets of Opera neutrinos created on our side (in $g_{\mu\nu}$) to jump to the conjugate side (in $\bar{g}_{\mu\nu}$) to allow a kind of oscillation between the two and eventually a detection on our side (in $g_{\mu\nu}$). Fortunately such jumps are actually expected in the Dark Gravity theory [6] through discontinuities of the gravitationnal field which are most often encountered inside matter.

To explore in the most simple way what kind of magnitudes of superluminal effects are to be expected in such framework, let us assume an external fictitious observer living in the flat non dynamical background. For this observer a photon or any ultrarelativistic particle propagating in the Schwarzschild $g_{\mu\nu}$ field generated by a nearby spherical Mass M on our side follows its geodesics hence:

$$0 = (1-2GM/rc^2) dt^2 - (1+2GM/rc^2) (d\sigma^2 \equiv dx^2 + dy^2 + dz^2)$$

while another photon on the other side follows the geodesics of $\bar{g}_{\mu\nu}$

$$0 = (1+2GM/rc^2) dt^2 - (1-2GM/rc^2) (d\sigma^2 \equiv dx^2 + dy^2 + dz^2)$$

where we have assumed $\bar{g}_{\mu\nu} = \eta_{\mu\rho} \eta_{\nu\lambda} g^{\rho\lambda}$ to hold (hence the conjugate metrics elements are inverse to each other) in the coordinate system where the conjugate fields have the isotropic form (not the standard one) and retained only some Post Newtonian approximated metric elements because we assume that we are in a very weak gravitationnal field. Hence from the $\eta_{\mu\rho}$ point of view the PN approximated speed of light in $g_{\mu\nu}$ is $d\sigma/dt = 1-2GM/rc^2$ (resp $d\sigma/dt = 1+2GM/rc^2$ in $\bar{g}_{\mu\nu}$) so the speed of light in the conjugate metric is $1+4GM/rc^2$ times faster. This ratio between these two observables would be the same if we had considered an observer in either of the two conjugate metrics instead of the observer linked to $\eta_{\mu\rho}$. So this is the result we needed: the ultrarelativistic neutrinos propagating in the conjugate side will appear to propagate at a speed $1+4GM/rc^2$ times faster than c ! However, as we shall explain in the next section, we expect the neutrinos detected at Gran Sasso to have propagated alternatively at the speed of light c on our side and $1+4GM/rc^2$ faster than the speed of light on the conjugate side, hence most probably a mean speed $1+2GM/rc^2$ faster than the speed of light. In other words we expect $\delta v_\nu / c = 2GM/rc^2$.

In order to compare the predicted $2GM/rc^2$ and the observed $\delta v_\nu$ we now need to know what is the dominant contribution to the adimentional potential $GM/rc^2$ on earth. It is neither the sun one, nor the even smaller earth one, but the potential due to our local cluster of galaxies well. The gravitational redshifting effects of such potentials [9] was even recently measured to be in good agreement with GR expectations for nearby clusters and the order of magnitude of the adimentional potential $GM/rc^2$ is generally a few $10^{-5}$. This is not surprising since we would expect $GM/rc^2 \sim (v/c)^2$ (this is just an order of magnitude) and we know the typical galaxy velocities relative to the CMB (our speed relative to the CMB is 600km/s) to be 1000 km/s hence typically $GM/rc^2 \sim 10^{-5}$ hence $2 GM/rc^2$ remarkably compatible with the $\delta v_\nu / c$ measured by Opera.

## III No conflict with SN1987 neutrinos

Discontinuities in the Dark Gravity [6] theory are considered to be genuine metric crossings hence switches allowing particles to jump from one metric to the conjugate one. Though this sector of the theory was only superficially explored a reminder of the motivation for field discontinuities in physics in general, their theoretical status and how they appear in the Dark Gravity theory and the best evidence we have for their existence (the Pioneer effect) will be given in forthcoming sections. But even if we had no sure quantitative description concerning the effect of such discontinuties on wave pakets crossing them, one would reasonably expect that, as for potential barriers in QM, parts of the wave paquets should reflect or instead be transmitted on both metrics, and this is all we need here. Such discontinuities are also expected to be concentrated in the vicinity of atoms, so it is only when they propagate a long distance inside matter that particles should encounter many of these discontinuities and consequently see their probabilities to jump to the other side become very significant and even to have their wave paquets equally splitted, oscillating between the two conjugate metrics. Thus for photons propagating in optical fibers it might also be interesting to check for superluminal effects with the same order of magnitude as in Opera or to compare the observed level of absorption of photons with theoretical expectations since the jumping of the photons to the conjugate metric also contributes to a disappearing effect.

Anyway, from our new understanding of the behaviour of neutrinos propagating inside matter and oscillating between the two metrics, we expect the neutrinos detected at Gran Sasso to have propagated alternatively at the speed of light c on our side and $1+4GM/rc^2$ faster than the speed of light on the conjugate side, hence hopefully a mean speed $1+2GM/rc^2$ faster than the speed of light. The context is very different for neutrinos emitted by SN1987 which might have oscillated inside matter before completely escaping the star, but for sure then could not anymore oscillate during most of their travel in vacuum to the earth, in which case we expect some wave paquets, having propagated mainly on our side to arrive on earth almost synchronized with the light from the SN while the huge delay (years) for the conjugate side wave paquets would prevent any correlation with the SN1987 event. Indeed the delay between neutrinos having propagated $1+4GM/rc^2$ faster than the speed of light on the conjugate side and the received flash of light would have been 8.3 years and we have heard about no neutrino detector already active in the late seventies. Therefore we find no conflict with the SN1987 neutrinos observation.

# IV Closed Timelike Curves and Causality

It is well known that signals propagating faster than c might lead to the possibility of Closed Timelike Curves hence causality violations. This is already a problem in GR where several field solutions exhibiting CTCs were explored. The very simple thought experiment usually treated in basic special relativity courses to explain how such CTCs might occur considers an exchange of signals propagating faster than c (let us say a tachyonic speed $C_T$ relative to the emitters of the signal) between me on earth and a distant astronaut going away at speed u relative to me.

In my reference frame R' the travel time on the way out is $\Delta t'_1 = \Delta x'_1/C_T$ while in the reference frame R of the distant astronaut the travel time on the way back is $\Delta t_2 = \Delta x_2/C_T$.

According to the distant astronaut $\Delta x_2$ is the distance $\Delta x'_1$ contracted by the relativistic factor $\gamma$ plus the distance $u \Delta t_2$ i went through during the propagation time of the signal on the way back: $\Delta x_2 = \Delta x'_1/\gamma + u \Delta t_2$ hence $\Delta t_2 = \Delta x'_1 / (\gamma.(C_T-u))$ and $\Delta x_2 = C_T \Delta x'_1 / (\gamma.(C_T-u))$. Replacing in the expression $\Delta t'_2 = \gamma (\Delta t_2 - u \Delta x_2/c^2)$, we find $\Delta t'_2 = (1 - u C_T / c^2) \Delta x'_1 / (C_T-u)$.

The total round trip duration for me is then $\Delta t'_1 + \Delta t'_2 = \Delta x'_1 (1/C_T + (1 - u C_T / c^2) / (C_T - u))$ negative if

$$C_T / c > (1+1/\gamma) c / u$$

which is the condition for the existence of a CTC in this case. For $u \ll c$, this is only possible for $C_T \gg 2c$ which we should consider seriously only in a strong gravitational field in our framework. But unless we are in the vicinity of a very compact object, we should always be in the case $C_T / c = 1 + \delta$ with $\delta \ll 1$ as in the case of Opera's neutrinos, and a CTC is only possible for $u/c = 1 - \varepsilon$, $1 \gg \delta^2 > 2\varepsilon$, thus my astronaut colleague should be ultrarelativistic relative to me.

Therefore it appears that though CTC are expected to occur, this will only be the case in extreme situations which might not be accessible experimentally. It is also usefull to recall that there are many open ways to be explored to hopefully avoid causality problems, one of which is the Novikov self-consistency principle if we admit that there is only one timeline accessible but assuming many alternative timelines accessible or many parallel histories in a single time (taking serious the many-worlds interpretation of quantum mechanics for instance) would also be a fascinating possibility.

# V Gravity and the Quantum: Strong Theoretical Motivations for DG Field Discontinuities

Genuine discontinuities are completely banned from modern physics. Indeed the derivation of all our fundamental interactions differential equations and conservation laws can only follow from the postulated actions invariance under various fundamental symmetries provided there are no field discontinuities. For example, the absence of discontinuities belong to the set of mandatory conditions for the Noether Theorem to be valid so even the local conservation of energy and momentum is not in principle granted anywhere we would encounter a field discontinuity. However, there are strong clues that at a fundamental level field discontinuities should be taken very serious. We know that an extremely enigmatic process, discontinuous and non local, the collapse of the wave-function, is one of the fundamental postulates of Quantum Mechanics and all modern physics of course must respect the rules of QM just because Nature was found to behave according these rules (even the non-local essence of the collapse is now very firmly established by many beautiful Quantum Optics experiments). It is important to realize that QM describes physical phenomena by two very distinct sets of rules. Let us stress this.

The first set of rules drives the continuous space-time evolution of various field solutions of the fundamental differential propagation equations of the fields (Klein-Gordon, Dirac ...) which can also be understood as local conservation equations and can also describe the interactions between all the fundamental fields once various local Gauge Symmetries are demanded.

The second set of rules was completely unexpected and very disturbing because it seemed to incredibly ignore all the beloved principles underlying the first set: these are the strange projection that mathematically describes the QM collapse and the Planck-Einstein relations, $E = h.f$, both completely unfamiliar to all the rest of physics. Both are discontinuous and non local in essence! Many physicists were indeed soon very dissatisfied with QM and Einstein himself believed that a more fundamental theory were to be found, also because QM is fundamentally undeterministic. This hope seems to be now completely given up just because we now know for sure that any such more fundamental theory underlying QM, a so called hidden variable theory, would have to be explicitely non local and most physicists prefer QM as it is (a set of rules that should not too much be taken serious thus a positivist interpretation rather than a realistic one) rather than trying to build a new framework with a set of explicitely non-local and discontinuous rules and principles drastically different from everything else we were used to think about seriously when constructing classical theories. Yet my conviction is that discontinuous fields and non local interactions are absolutely mandatory if one would really want to elucidate the origin of the as well discontinuous and non local rules of QM (and hopefully compute the value of the Planck constant h from more fundamental space-time parameters). So, from this point of view, it should actually be considered as an incredible advantage to have a new theoretical framework in which discontinuities are natural and necessary and not a drawback as most theoretical physicist would think nowadays.

The initial motivation for a theory such as Dark Gravity was not at all to stage a priori shocking new rules such as non local interactions and field discontinuities but the very constraining principles of the theory led to it and eventually this is unhoped-for. At a more fundamental level the two sets of rules we found in QM, one continuous and local and the other discontinuous and non local will hopefully emerge from the structure of the DG theory which admits both a sector of usual propagated interactions but also field discontinuities and a non local sector for gravity. This makes it a « dream » theory not only to unify QM and gravity but more importantly to really explain where the QM strange discrete and non local rules come from and derive them, a program i also started to explore in [6].

Why are field discontinuities naturally expected in DG ? Just because the theory follows from a new treatment and understanding of space-time discrete symmetries, and therefore its fundamental equations admit two time reversal conjugate solutions to describe the background (cosmological type solutions) for instance. These are

$$d\tau^2 = a^2(t) (dt^2 - d\sigma^2 \equiv dx^2 + dy^2 + dz^2)$$

while on the other side

$$d\tau^2 = a^{-2}(t) (dt^2 - d\sigma^2 \equiv dx^2 + dy^2 + dz^2)$$

(by the way $a^{-2}(t) = a^2(-t)$: the metrics are time reversal conjugate) and because time reversal can occur a priori anywhere, there is no reason why a single of these two solutions should be valid everywhere on our side of the universe i.e there is no reason why the solution on our side should be $a(t)$ (or $a^{-1}(t)$) everywhere and should be $a^{-1}(t)$ (resp $a(t)$) everywhere on the conjugate side. Instead, we naturally expect the universe to be divided in spatial zones where different solutions were chosen. For instance there might be an expanding solution on our side in the solar system (and the conjugate contracting solution on the conjugate side) replaced by a contracting solution outside the solar system (and the conjugate expanding solution on the conjugate side). Of course this implies a genuine discontinuity of the background field at the frontier between the two zones at which a genuine time reversal occurs and the conjugate solutions are exchanged.

# VI The Pioneer effect: Strong Observational Motivation for DG Field Discontinuities

Now what kind of signature could reveal the existence of such discontinuities ? Of course light is trivially not affected by conformal metrics ($a(t)$ or $a^{-1}(t)$ have no effect when $d\tau = 0$) therefore light propagating on one side of the universe is not sensitive to discontinuities of the background but of course its wavelength should be shifted when jumping from one side of the universe to the other by the potential difference $2GM/rc^2$ implied between the emitter side where the feeled gravitational potentials is $-GM/rc^2$ and the receiver side where the feeled gravitational potential is the opposite $GM/rc^2$. By the way, we can mention that we suspected in [6] that this gravitational wavelength shift of CMB photons transferred from one side to the conjugate one through discontinuities might have contributed to or even been responsible for most CMB fluctuations, since the order of magnitude of such redshifts corresponds to typical temperature fluctuations in the CMB (a few $10^{-5}$).

Now what about massive particles or atomic clocks. Imagine two identical clocks exchanging light signals from both sides of the border line where you have a discontinuity (again such signals are unperturbed when crossing the discontinuity). Then they can compare the speed of time in the two zones, one zone where the background metric field element is $a(t)$ and the other where it is $a^{-1}(t)$. The frequency shift one clock will see comparing the frequency of the other clock with its own can be computed as in [10], because on one side:

$$d\tau^2 = a^{-2}(t) (dt^2 - d\sigma^2) \quad (1)$$

which yields for a clock at rest ($d\sigma^2 = 0$) there (suppose on earth):

$$dt_{Earthclock} = d\tau . a(t)$$

while for a clock at rest on the other side (suppose Pioneer is there)

$$d\tau^2 = a^2(t) (dt^2 - d\sigma^2) \quad (2)$$

$$\Longrightarrow$$

$$dt_{Pioneerclock} = d\tau / a(t)$$

This obviously implies that the Pioneer clock frequency will drift in time as compared to our earth clock. But then shouldn't the period of the Pioneer clock have been suddenly rescaled by a huge $a^2(t)$ factor when the spacecraft crossed the discontinuity. Not necessarily if at a more recent time $t_0$ the conjugate background fields started an evolution of the kind

$$d\tau^2 = a^{-2}(t_0) \, a^2(t)/a^2(t_0) \, (dt^2 - d\sigma^2) \quad (3) \text{ our side}$$

$$d\tau^2 = a^2(t_0) \, a^{-2}(t)/a^{-2}(t_0) \, (dt^2 - d\sigma^2) \quad (4) \text{ on the conjugate side}$$

rather than (1) and (2) and that this occured in a spacial zone where Pioneer was, while on earth we remained in a zone where (1) and (2) was valid.
Then for our Pioneer and Earth clocks in two zones exchanging the roles of metric conjugate solutions (1) and (3).

from (1)    $dt_{Earthclock} = (a(t)/a(t_0))d\tau \, a(t_0)$
from (3)    $dt_{Pioneerclock} = (a(t_0)/a(t))d\tau \, a(t_0)$

Fortunately, the Pioneer effect is instructive: it tells us that clocks periods are not instantaneously rescaled by a huge $a^2(t)$ scale factor that would have followed from (1) and (2) when crossing discontinuities but rather the much smaller rescaling $dt_{Earthclock} = dt_{Pioneerclock} \, a^2(t)/a^2(t_0)$ following from (1) and (3) which absolute effect, $f_{Pioneer} = f_{earth} \, a^2(t)/a^2(t_0)$, the frequency resolution of the electronics was insufficient to detect, unlike fortunately the frequency drift, genuine deceleration of Pioneer clocks relative to our earth clocks, the anomalous:

$$\frac{\dot{f}}{f} = 2 H_0 = 4.8 \, 10^{-18} s^{-1}$$

where we have used $H_0 = \frac{\dot{a}}{a}$ expression for the Hubble parameter in conformal coordinates which is easy to check.

This result is remarkably compatible with the one that was measured, $\frac{\dot{f}}{f} = (5.6 \pm 0.9) \, 10^{-18} / s$, when analysing the Pioneer spacecraft radiowaves. Therefore, we are tempted to conclude that there must have been a discrete jump from $a(t)/a(t_0)$ to $a(t_0)/a(t)$ of the background field between us and Pioneer so that the effect could only start to be seen after the crossing of this frontier by the spacecraft. Within the error bars the jump (see the steep rise up of the effect in [16]) could not have been better evidenced than it was around 15 AU in 1983 by Pioneer 11. This extraordinary evidence, the perfect expected signature for a background discontinuity is the fact that convinced me that such discontinuities, a priori naturally expected in DG are actually real and have observational consequences. Indeed, up to now nothing else appart from this kind of very particular shift in time can account for the Pioneer anomaly without conflicting with many other precision tests of gravity [10] in the solar system and the main possible systematical effect, an anisotropic radiation from the spacecraft, could only account for a small fraction of the effect.

However, trusting the description of the communication systems given in [16] we realize that if the emitter on board of the Pioneers was PL Locked on the signal received by the spacecrafts Antennae, in principle, the pure frequency multiplication performed there between downlink and uplink should not have been sensitive to the actual background metric felt by the electronic on board. This issue is adressed in our new version of [6].

The Pioneer effect also tells us that because all clocks in the universe did not have exactly the same history, this could eventually lead to huge redshift anomalies even between clocks relatively closed to each other i.e. not at cosmological distances from each other. Because such anomalies was not observed we must conclude that on the long term, particularly on cosmological times, all clocks have experienced almost the same relative elapsed time in regime $a(t)$ and $a^{-1}(t)$ on average. This in turn is only possible provided discontinuities such as the one responsible for the Pioneer effect are themselves moving, drifting everywhere probably cyclically. Anyway, we are now well motivated to start to explore many other possible phenomenological consequences of such discontinuities, such as an instantaneous boost of massive particles crossing the potential barrier implied by $a(t)/a(t_0)$ and shall now show that such impressive signatures do exist in LENR phenomena.

## VII Field Discontinuities explain LENR Phenomena

All searchers in the field of LENR phenomena are probably aware that LENR is not only associated with:

A. Large extra heat (not possibly of chemical origin) with very low levels of nuclear radiations (alpha, beta, gamma, neutrons) as compared to what would be expected from nuclear processes producing the same amount of energy.

B. Transmutations and isotopic anomalies.

But also very often with:

C. Observation of a new category of incredible objects which behaviour seems almost impossible to understand without postulating new physics (for instance caterpillar traces left by micron sized magnetic and radiating objects able to fly meters away from their source [12], to go through dense materials, to explode and release much energy in them[13], etc … ) objects which were discovered by many scientists independently (Matsumoto, Dash et al., Shoulders, Lewis, Savvatimova, Urutskoev et al*.* , Ivoilov and other groups) in many kind of experiments involving macro or micro electric discharges and independently named Evos, EVs, Ectons, Plasmoids, Ufos (for instance in Tokamaks or at the LHC), Leptonic Monopoles, Charged Clusters, Nucleon Clusters, Micro Ball Lightning, etc ...

In my opinion, any idea proposed to explain A or B but neglecting C is almost certainly wrong because it is unlikely that two kinds of very different new ideas would be needed, one to explain C and another to explain A and B, while the detections of the two kind of effects are clearly related. Indeed there is even an annual conference called Russian Conference on Cold Nuclear Transmutation and Ball-Lightning (RCCNT&BL) and there also have often been presentations on Ball Lightning at the ICCF conferences.

On the other hand if you are able to provide an explanation for C which at the same times clarifies A and B, ... Bingo!

Good references to start to gather a list of typical properties of these objects are [11] (see references therein) [12] [13] and i personally consider, following Lewis [11] that these objects can exist with very variable sizes and lifetimes and are all of the same nature as the much bigger Ball Lightning sometimes observed in thunderstorms so from now on i will generically call them micro ball lightning, or mbl. Their common source is most probably always an electric discharge including micro-discharges near metal surfaces in simple electrolysis experiments or in experiments where these discharges can result from the metal surface being submitted to mechanical, thermal or EM pulse shocks.

The most obvious of the mbls properties is the density of particles inside them comparable to the one in condensed matter in some cases [13] but as elevated as nuclear densities in some others [15] … most probably this density is determined by the medium the mbl is propagating through (gaz, liquid, solid) and might afterward be compressed up to nuclear densities inside the mbl [15]). Knowing that the temperature inside is at least of thousand degrees and may be up to the hundred millions of degrees needed to trigger nuclear processes, the pressure inside and the energy densities the mbls are able to carry can be phenomenal and be delivered to the environnment either during their lifetimes or in an ultimate explosion.

As well as for macroscopic Ball Lightning the challenge is thus to find a mechanism able to confine this huge energy density and resist the pressure during the whole lifetime of for instance a 6 micron sized mbl (between 10 and 0.01 microseconds) [14] and explain how such a macroscopic collection of a huge number of particles can behave as a single object leaving a well defined track in nuclear emulsions or boreholes in matter.

The stability problem is of course even worse if you take serious the results of various searchers (K. Shoulders[13], Rambaut[14]) strongly suggesting that the mbls also carry a huge electric charge because you then have to resist the corresponding electrostatic repulsion between an incredible concentration of charges of the same sign. On the other hand, L. Urutskoev [12] demonstrated that these mbls carry a magnetic charge (can be trapped in iron and affect Mossbauer spectra) thus suggested to identify them as leptonic magnetic monopoles, objects theoretically predicted by G. Lochak [12], but also argued that mbls could not carry an electric charge otherwise these could not be able to pass through two meters of atmospheric air and two layers of black paper as they did and yet mbls must be charged as nuclear emulsions are insensitive to neutrons.

These seemingly contradictory statements can be reconciled once you understand the mbl not only as a collection of particles but as a collection surrounded by a huge spherical discontinuous gravitational potential which can accelerate in a centripetal way all massive particles encountered up to an energy proportional to their mass and then trap them inside, resisting both pressure and electrostatic repulsion between particles of the same charge trapped in the volume delimited by the discontinuity. Then how is an mbl able to propagate such a long distance in a dense medium if it has a charge? The spherical discontinuity can fuse, evaporate or even turn all the material encountered by the discontinuity into a plasma often without apparently slowing down the mbl. Many searchers also noticed that the tracks left by mbls show sharp angle turns manifesting huge accelerations as if the mbl as a whole did not have any inertia. This can be understood if the discontinuity which defines the periphery of the object moves following its own laws and carries with it all the matter content inside it in the same way whatever the mass and interactions of this content. For instance if the discontinuity is located along an isopotential electrostatic surface of the cluster of particles inside, any motion of the cluster charge distribution inside will also globally shift the discontinuity and this in turn will carry the cluster trapped by the huge discontinuous potential well. The motion of the mbl through any medium thus would appear to be self-sustained and not resisted as far as the new neutral matter injected in the mbl by the encountered medium, as seems to be often the case, would not produce huge perturbation of the charge distribution inside the mbl. However suppose for instance that an external electrical field is applied : this will accelerate much more efficiently the particles with a big ratio charge/mass inside the cluster hence the electrons and these in turn will drive the electrostatic isopotential where the discontinuity is sitting hence it is the whole mbl that will be accelerated as efficiently as each single electron in the electric field! Therefore phenomenal accelerations of mbls are possible and for instance mbls can also describe circles at high cyclotron frequencies in a magnetic field as was also often observed [11]. Such kind of observations might have created the illusion that the mbls manifesting so weak inertia (huge q/m ratios) were merely electron clusters.

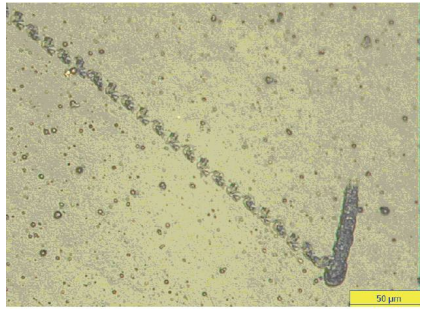
Sharp angle turn track on X-ray film.

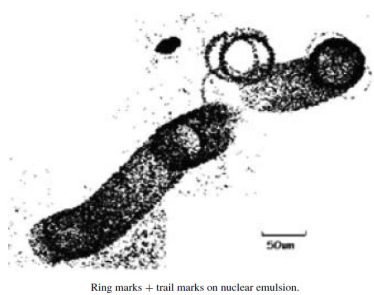
Ring marks + trail marks on nuclear emulsion.

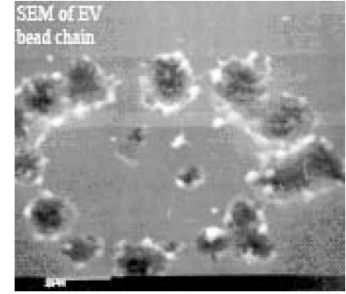
**Figure 4.** The scale is 25 $\mu$m.

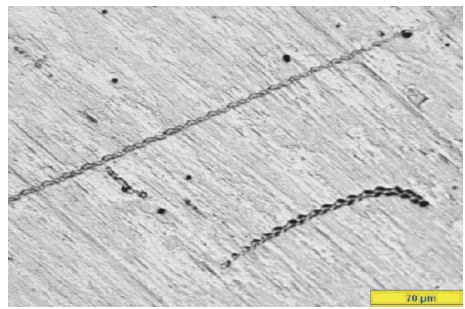

Sharp angles, rings and strange tracks
(Savvatimova, Matsumoto, Shoulders, Savvatimova)

The discontinuity is of course one possible source of the particles kinetic energies and hence temperature inside the mbl but if the temperature is high enough, of course nuclear processes can be triggered and another kind of potential energy, nuclear rather than gravitational, be liberated. But how could the energy escape out of the mbl and be measured as heat outside if the energetic particles are all trapped inside? Again the answer is simple: the kind of gravitational potential barrier implied by a discontinuity of the background field has no effect on massless particles (conformal metric), so any photon can cross it and escape (hence the name Ball Lightning). Thus the radiative cooling of the mbls can take place efficiently and these can heat the environment in that way provided their lifetime is long enough.

How might the mbls acquire and be stabilized with an electric charge ? Let us consider for instance a spherical discontinuity accelerating each nucleon of the matter around, to 20 keV ( $m_n$ . $2GM/rc^2$). This also means that any nucleon inside the mbl with energy below this 20keV threshold cannot overcome this potential barrier and escape the mbl. The electrons (2000 times less massive) will be accelerated to 10 eV only by the same discontinuity and any electron with energy below 10eV will not be able to escape the mbl for the same reasons. Now inside the mbl any interaction between the cold electrons and very hot nucleons will likely boost the electrons at an energy much above the 10eV threshold. Thus eventually the electrons tend to be ejected out of the mbl while the nucleons are trapped much more efficiently because the potential barrier is much higher for them. This would result in a huge positive electric charge for the mbl if it was not counter-balanced by the electrostatic attraction of the escaping electrons by the protons inside the mbl. Eventually at equilibrium a stabilized mbl would have a core positive charge in a somewhat more extended cloud of electrons like a kind of huge atom which would radiatively cool very fast down to the temperature at which both electrons and nucleons would be trapped by the surrounding discontinuous potential barrier. This huge atom is by the way also an interesting picture to hopefully later understand the atomic structure in a new way i.e where the most intriguing QM rules (the Planck-Einstein quantization rules) come from !

However this pictures a globally neutral mbl while we have seen why the mbl behaviour in an EM field is strongly suggesting that its discontinuity is sitting along an electrostatic isopotential. It is therefore tempting to suggest that any potential whether electrostatic or gravitational reaching a given threshold (this implies either a concentration of charges or mass) might trigger the apparition of a discontinuity. For instance an electric discharge impact might generate a very short-lived concentration of charges of the same sign which is of course electrostatically very unstable and should disperse very fast if a discontinuous potential suddenly appearing did not trap them, stabilizing the object for a much longer time. Such discontinuity would appear because of the electrostatic potential reaching the required threshold in the vicinity of the concentration of charged particles. The mbl would therefore be stabilized as long as it is able to keep the charge that gave birth to it.

Can we really test the presence of discontinuities at work in mbls specifically the expected temperature discontinuity between inside and outside the mbl? Let us cite the ground breaking result obtained by Shoulders after analysing boreholes left by mbls [13]: «The borehole is fairly clean for a process that is capable of fluidizing a material with a melting point of 2,600 degrees centigrade and projecting it to an unholy velocity. In fact, when a special test is set up to determine the thermal gradient at the edge of the borehole, one comes to an astounding conclusion: either a gradient of over 26,000 degrees centigrade per micrometer exists here, or this is a non-thermal process!» : can we imagine a better signature for a discontinuity ?

The comparision with inertial fusion confinement is quite instructive: often the compressed fuel target has a typical diameter of 200 microns, a density 1000 times (d=1000) that in condensed matter (d=1), and a temperature of 10 keV and for these parameters the energy confinement time $\tau$ is several $10^{-11}$ seconds. Then if the Lawson criterium is respected for d = 1000 and $\tau \sim 10^{-11}$ seconds, this should also be the case for d=1, $\tau \sim 10^{-8}$ seconds at the same temperature and for the same DT nuclear fusion process and for d=1, $\tau \sim 10^{-5}$ seconds still at 10 keV for DD nuclear fusion [17].

These numbers seem to imply that at least in those cold fusion experiments involving Deuterium and producing Helium, a significant amount of energy might be produced by nuclear fusion processes in mbls with d near 1 and a temperature approaching 10keV. Such hot mbls could be the result of discontinuities accelerating nucleons to keV energies hence may be a link between $a(t)/a(t_0)$ and $m_n . 2GM/rc^2 = m_n . \delta v_\nu / v_\nu \sim 20$ keV. However, the fact that no significant rates of high energy neutrons has been observed in most of these experiments certainly implies that this picture of a very hot and d near 1 mbl is not correct and that the nucleons entering the mbl and crossing the discontinuity are accelerated by the unknown potential barrier $a(t)/a(t_0)$ to much less than keV energies.

Then there is still the possibility that the energy eventually radiated in the environment by the mbl is not at all of nuclear origin but just originates from the gravitationnal energy gained by the nucleons when crossing the mbl discontinuity but then how could we interpret the detection of Helium in some of these experiments and the common isotopic anomalies encountered in those LENR experiments with Hydrogen rather than Deuterium ?

It is interesting to notice that even from the hot mbl that we have already excluded (an mbl surrounded by a discontinuity accelerating nucleons to 20 keV hence able to trigger nuclear processes and MeV radiations) most high energy particles would be thermalized before escaping the mbl. Indeed, the main difference between the conditions inside such an mbl and in the sun or Tokamak plasma is the much higher density (d=1) hence pressure which results in a mean free path, for all particles except for neutrons, completely neglible relative to the size of the mbl: therefore any MeV particle released by a nuclear process inside the mbl should be almost instantaneously thermalized to a mean energy below 20 keV inside the mbl well before reaching the outside frontier of the mbl i.e the discontinuity. In other words, no alpha nor beta nor gamma particles can escape the mbl in the MeV range except a very small fraction produced very near the surface of the mbl. However this would not prevent high energy neutrons to be radiated. The very small rates of high energy neutrons in LENR experiments, therefore again favours mbls with much lower mean temperatures than those that would trigger nuclear fusion processes in such a way that they would unavoidably produce large rates of high energy neutrons. The question therefore is: how could mbl help trigger nuclear processes (Helium and isotopic anomalies) without the associate high energy neutrons ?

The case of hydrogen (instead of Deuterium) LENR experiments in which large isotopic anomalies are commonly reported, strongly suggests that the mbl might be better understood as a very high pressure rather than a high temperature object. The theory of neutron stars formation is instructive: for instance it was computed that between $8.10^6$ and $1.5\ 10^9$ grams per cm$^3$, the very dense matter is best decribed as a crystal of the most stable nuclei in such conditions embedded in a liquid of electrons. These Nuclei are mostly $^{62}$Ni, $^{64}$Ni which are also found to be produced with very unnaturally high proportions in the Rossi reactor. Perhaps the implosion of mbls (centripetal running of the discontinuity) might lead to such extreme densities inside them. Then the nuclear fusion processes are not made possible by huge temperatures i.e kinetic energies of the particles inside mbls allowing the nuclei to overcome the Coulomb potential barrier, but rather by the screening effect of the electrons (in between the nuclei to be fused) in this particular state of matter compressed by the imploding mbl. One would expect that the fusion processes could only start when the pressure and density are so huge inside such an mbl that we are approaching the picture of the nucleon ball (just as the ones discovered and described in [15]). The fusion reactions are then many body processes and at such densities the mean free path of a MeV neutron would be so small that even a nanometer sized mbl would efficiently slow down any such neutron produced inside it before it could reach the mbl surface and escape. Indeed, the mean free path for MeV neutrons in hydrogen at atmospheric pressure with the cross section around $\sigma = 2\ 10^{-24}$ cm$^2$ and density d

= 3 $10^{19}$ molecules/cm$^3$ is λ = 1/σd ~170 m, so if we start from a 10 microns radius mbl containing gases and imploding to a nanometer radius mbl, the density inside d is expected to increase by a factor 2 $10^{4\times3}$ (factor two because of the dissociation of H$_2$) hence we expect a final λ near the Angstrom so that the MeV neutron produced at the center of such compressed mbl will scatter more than 10 times before escaping the mbl. Knowing that a 2 MeV Neutron looses on average half of its energy after 15 collisions in Hydrogen [20] and that the scattering cross section increases when the neutron energy decreases, the slowing down of the neutrons is already effective in such a nanometer sized mbl and this would have been the case even earlier in the implosion process, had we started from a micron sized mbl with already the density of condensed matter. Eventually the nuclear energy is only produced at compression levels allowing the fusion of the nuclei, but then the density is so high that any neutron should be thermalized before escaping the mbl (if its remaining energy allows it to cross the discontinuity), so most of the nuclear energy is expected to escape the mbl as relatively soft EM radiations only.

But one should also keep in mind that mbl can heat their environment even if their temperature and density inside are too low to trigger nuclear processes.

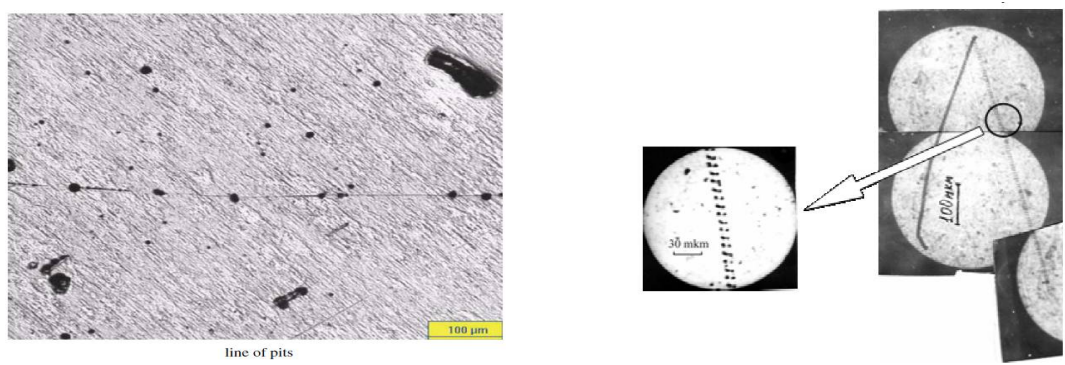

Line of pits and caterpillar tracks
(Savvatimova, Urutskoev )

Eventually let us not forget that discontinuites in DG are connecting the two sides of the universe. This is why the material content of the mbl might oscillate between our side of the universe and the conjugate side (the antimatter universe) via the peripheral discontinuity of the mbl so that the mbl may have an alternating luminosity from one side (the observer side i.e our side) point of view, hence leave strange caterpillar or dotted line traces in emulsions as described in [12] for instance. At last, the matter inside the mbl might manifest a ferromagnetic or even ferrimagnetic behaviour depending on how particles align their spin inside so it's not completely surprising that such objects can be trapped in ferromagnetic materials [12].

## VIII Thermodynamic analysis of an mbl

Let's start with an mbl with a 1 micrometer radius, a few thousand degrees at most (as suggested by observations of larger bl in thunderstorms) implying that the discontinuity potential barrier is not much larger than 1eV for nucleons and 0.5 meV for the electrons, and the typical nuclei and electrons densities of condensed matter. As we already explained the mbl has a charge because of an initial random excess (a fluctuation in the transient plasma of a powerful electric discharge) of electrons (resp nuclei) over nuclei (resp electrons). Let us further assume that the mbl is in vacuum hence no new fresh matter can feed it! Because the stability of an mbl is understood to be determined by its ability to retain its charge, negatively charged mbl should be very unstable and short lived because the 0.5meV potential barrier is not strong enough to trap the excess of electrons electrostatically repelling each other, all the more since the electrons can also be ejected from the mbl by the much hotter nuclei (initially heated to 1eV). As a result of the mbl loosing its charge, its electrostatic potential drops resulting in the very fast centripetal running of the discontinuity : that is a genuine implosion of the mbl! As a result, the massive content of the mbl, the ions, is compressed to very high pressure in a very short time. If the pressure work W is delivered to the plasma so fast that we can neglect the energy radiated away given the temperature T and surface S of the mbl ($P_{rad}$ = σ S T$^4$), the compression is almost adiabatic and the plasma is heated according T ≥ Trev ~ Const/V$^{γ-1}$ where Trev is the expected temperature law for an adiabatic reversible process i.e if the internal pressure equals the equivalent of an « external pressure applied by the discontinuity » at any time during the process which is not unreasonable even for a fast collapse for such a small object. The heating proceeds up to temperatures allowing the nuclei to overcome the 1eV potential barrier and escape the mbl: in other words, as a result of loosing its charge excess, the mbl also looses its mass in the implosion process so that until its complete disappearing (such a progressive « evaporation » of an mbl has probably been captured on nuclear emulsions in [12] where the width of the strange tracks is reduced in proportion to traveled distance) neither its temperature (particles hotter than threshold escape) nor its pressure (matter escapes) will raise to unholy values.

The evolution might be very different if the mbl is positively charged because the excess charge is now carried by the ions which are trapped very efficiently by a 1eV potentiel barrier, because the content of a mbl in vacuum can only cool by radiative losses and losses due to the hottest ions escaping the 1eV barrier the proportion of which is expected to drop as the temperature falls. Thus, in perfect vacuum such mbl might be perfectly stable if the mbl content keeps cold enough, however if there is matter around, because of its positive charge it must try to recover neutrality by attracting free electrons around and doing so, loosing its net charge, its electrostatic potential drops as in the negative mbl case, resulting again in the centripetal running of the discontinuity. What is new and interesting is that now such implosion might be very slow, only depending on the availability of electrons around.

According to [12] it is indeed possible to trap what the authors called monopoles (our mbl) in ferromagnetic materials for days.
What we know for sure is that there is a maximum temperature of a few thousand degrees that the mbl content cannot exceed for a 1eV potential discontinuity, so any implosion of our mbl that would be too fast to allow the mbl to radiate its energy would result in exactly the same implosion scenario and fate as already described for a negative mbl.

But now, let us investigate the case of the very slow implosion of an mbl as it progressively recovers neutrality. A fascinating new phenomenology opens as nothing now prevents our mbl, just as a white dwarf, to be compressed to huge densities ($10^6$ higher than condensed matter ones) and pressure of the ion gas as its size is divided by 100, even greater pressure of the electron fermi gas, while remaining a cool object, in equilibrium with its environment (still assuming a quite well isolated mbl with only a small input rate of electrons for a slow come back to neutrality), thanks to its ability to radiate all the energy the compression work of the discontinuity gives to the ions as it goes along.

Actually any isolated white dwarf will also cool down to the temperature of its environment, i.e. may be down to fractions of a Kelvin over a time much longer than the age of the universe, and then become what has been called a black dwarf. Our mbl, just because of its much bigger ratio Surface/Volume can of course cool down in a fraction of a second and much faster than it is heated by the compression work of the discontinuity during its very slow collapse. Eventually a huge compression work is converted to heat over a long time in a LENR experiment. The next step will be the triggering of many body reactions involving electrons and nuclei, if the slow collapse goes on at higher pressures and densities. As the pressure gets higher and higher, the compressure work PdV can be huge even for a small dV which may result in a too fast increase of temperature to be compensated by radiation losses hence resulting again in a loss of mass (ions above 1eV) and fast « evaporation » and disapearing of the mbl. Eventually the loss of mass should also be unavoidable when the ions in turn will start to behave as a degenerate fermi gas as the ions will populate higher and higher energy states. But at intermediate densities and pressures, may be up to ten times that of the white dwarf, the picture of a ion gas with density near $10^7$ grams per $cm^3$, but as cold as 300K is fascinating! The de Broglie wavelength of the ions is then two orders of magnitude greater than their average separation and manybody interactions should lead the game.